\documentclass[12pt]{TD-CJS}

\usepackage{latexsym}
\usepackage{amsmath}
\usepackage{amsfonts}
\usepackage{amssymb}
\usepackage{psfrag}
\usepackage{graphicx}

\usepackage[flushleft]{threeparttable}

\usepackage{multirow}
\usepackage{bm}

\begin{document}


\rhauthor{Safari, Altman and Leroux}
\copyrightline{Statistical Society of Canada}
\Frenchcopyrightline{Soci\'et\'e statistique du Canada}

\renewcommand{\eqref}[1]{(\ref{#1})}
\newcommand{\mb}[1]{\mathbf{#1}}
\newcommand{\mbb}[1]{\mathbb{#1}}
\newcommand{\mt}[1]{\mathrm{#1}}
\newcommand{\rv}{random variable}

\title[]{Parameter-driven models for time series of count data}
\author{Abdollah Safari\authorref{1}\thanksref{}}
\author{Rachel MacKay Altman\authorref{1}}
\author{Brian Leroux\authorref{2}}

\startabstract{%
\begin{abstract}
\abstractsection{}
\abstractsection{Abstract}
This paper considers a general class of parameter-driven models for time series of counts. A comprehensive simulation study is conducted to evaluate the accuracy and efficiency of three estimators: the maximum likelihood estimators of the generalized linear model, 2-state finite mixture model, and 2-state hidden Markov model. Standard errors for these estimators are derived. Our results show that except in extreme cases, the maximum likelihood estimator of the generalized linear model is an efficient, consistent and robust estimator with a well-behaved estimated standard error. The maximum likelihood estimator of the 2-state hidden Markov model is appropriate only when the true model is extreme relative to the generalized linear model. Our results are applied to problems concerning polio incidence and daily numbers of epileptic seizures.
\end{abstract}}
\makechaptertitle


\section{Introduction}

\cite{Cox:1981} defines two classes of models for time series data: parameter-driven models (PDMs) and observation-driven models (ODMs). In an ODM, autocorrelation is introduced through the dependence of the conditional expectation of the current observation on the past observations (e.g. an AR(1)), whereas a PDM uses a set of latent variables to explain the autocorrelation. A key feature of PDMs is that the observed data are assumed to be independent given the values of the latent variables. The class of PDMs includes moving average models, hidden Markov models (HMMs), generalized linear mixed models (GLMMs), and hierarchical generalized linear models (see \citealt{Lee:Neld:1996}). ODMs and PDMs have different properties and usages. For example, in the case of ODMs, evaluating the likelihood and estimating the model parameters are typically straightforward, and, for this reason, ODMs are often used for forecasting. On the other hand, the interpretation of PDMs is usually simpler than that of ODMs. Specifically, on the marginal level, properties of PDMs are often easier to establish than those of ODMs, especially when the data are non-normally distributed. PDMs thus provide a convenient way of modeling overdispersion and autocorrelation. However, PDMs tend to have more complicated likelihoods, and hence estimation of these models is challenging. 

In this paper, we focus solely on the study of PDMs for time series of count data. In particular, we use the setting of \cite{Zeger:1988} where, conditional on the latent variables, the observations are Poisson distributed with log mean specified by a linear function of predictors and latent variables. \cite{Zeger:1988} introduces a consistent quasi-maximum-likelihood estimator (QMLE) of the regression coefficients in such models. Others have used this approach in different areas of application (e.g. Campbell 1994, Brannas \& Johansson 1994, Albert et al 1994, and McShane et al 1997). Davis, Dunsmuir and Wang 2000 (henceforth called DDW) suggest estimating the regression coefficients using the maximum likelihood estimator (MLE) of a Poisson generalized linear model (GLM). They prove that, although this estimator is based on a misspecified model, it is consistent for the true regression coefficients. They also derive its asymptotic covariance matrix. The GLM-based estimator has advantages like simplicity and robustness. However, to the best of our knowledge, its efficiency has not been studied. Similarly, the consistency of estimators of other PDMs is typically the only property studied (e.g. \citealt{Davis:Wu:2009} and \citealt{Neuh:McCul:Boylan:2013}).

Our goal in the present paper is to investigate the efficiency of the GLM estimator relative to that of other simple estimators. In particular, we evaluate the standard errors (SEs) of three estimators of the regression coefficients of Poisson PDMs: the GLM estimator, finite mixture model (FMM) estimator, and hidden Markov model (HMM) estimator. These estimators, unlike MLEs of some models, are easy to compute. In addition, we derive estimate of the asymptotic covariance matrices for these estimators.

The remainder of the paper is organized as follows. In \S \ref{Model specification} we specify the PDM of \cite{Zeger:1988} and present some special cases that we consider as a basis for estimating the regression parameters of this model. In \S \ref{estimate SE}, we derive asymptotic SEs of the estimators. In \S \ref{Sample SD}, we illustrate the performance of the estimators and their standard errors for certain choices of the true model. In \S \ref{Morestudies}, we present results on the relative performances of the estimators for a broad class of true models and provide some general rules for efficient estimation of regression coefficients in the context of time series of counts. In \S \ref{application} we present applications of our results to problems concerning polio incidence (in \citealt{Zeger:1988} \& DDW) and daily numbers of epileptic seizures. We conclude with a discussion in \S \ref{discussion}.

\section{Model specification and estimation}
\label{Model specification}

Let $Y = (Y_1, \dots, Y_n)$ be the observed counts, and let $\alpha = (\alpha_1, \dots, \alpha_n)$ be the latent variables. We assume, as for all PDMs, that $\{Y_t\}$ are independent given $\{\alpha_t\}$. Following \cite{Zeger:1988}, we model $Y_t \mid \alpha_t$ as having a Poisson distribution with
\begin{equation}
\label{MainModel}
\log \left( E\left[ Y_t \mid \alpha_t \right] \right) = X_t ' \beta + \alpha_t
\end{equation}
where $X_t$ is a d-dimensional vector of covariates at time $t$ and $\{\alpha_t\}$ is a general stationary process with $E\left[\exp\left(\alpha_t\right)\right]=~1$, $Var\left(\exp(\alpha_t)\right) = \sigma_{\alpha}^2$, $Cov\left(\exp(\alpha_t),\exp(\alpha_s)\right)=\gamma_{t-s}$, and $Corr(\exp(\alpha_t),\exp(\alpha_s))=\rho_{t-s}$, where $t \geq s$. We assume that the first entry of $X_t$ is a 1, i.e. that the first entry of $\beta$ is the intercept.

One way to understand the implications of the assumed (conditional) model on the resulting distribution for the observed data is to examine the marginal moments. In Zeger's model, the effect of the covariates on the marginal moments and the mean-variance and mean-covariance relationships are easy to determine. In particular, the first and second marginal moments are: 
\begin{eqnarray}
\label{Mean}
\mu_t \equiv E\left( Y_t \right) &=& \exp\left( X_t'\beta \right) \\
\label{Var}
Var\left( Y_t \right) &=&  \mu_t + \mu_t^2 \sigma_{\alpha}^2 \\
\label{Cov}
Cov(Y_s, Y_t) &=& \mu_s \mu_t \gamma_{t-s}, ~~ t \geq s \\
\label{Corr}
Corr(Y_s, Y_t) &=& \frac{\mu_s \mu_t \gamma_{t-s}}{\sqrt{\left(\mu_s + \mu_s^2 \sigma_{\alpha}^2\right)\left(\mu_t + \mu_t^2 \sigma_{\alpha}^2\right)}}, ~~ t \geq s 
\end{eqnarray}
Equation \eqref{Var} shows that $\mu_t^2 \sigma_{\alpha}^2$ is the extra-Poisson variation and is a function of $\sigma_{\alpha}^2$. In addition, from \eqref{Cov}, we can show that $\mid Corr(Y_s, Y_t) \mid \leq \mid \rho (\epsilon_s, \epsilon_t) \mid$, regardless of the distribution of the latent process (\citealt{Davis:2000}). Consequently, detecting and analyzing the latent process based on the observed data can be challenging.

Zeger's model is quite general, and, for this reason, includes many common models for time series of counts. We next discuss some special cases and explain how they can be used as a basis for estimating the regression parameters in \eqref{MainModel}.

\subsection{Poisson GLM}

If we set $\alpha_t \equiv 0$ in \eqref{MainModel}, then $\{Y_t\}$ are treated as independent and as following a Poisson GLM. We call the maximizer of the associated likelihood function the GLM estimator. DDW suggest using this (d-dimensional) estimator to estimate the parameters in Zeger's model and show that it is consistent for the regression parameters. However, the efficiency of this estimator is questionable since it may not use all of the information in the autocorrelation and overdispersion in the observations.

\subsection{Poisson FMM}
Since Poisson GLMs can't capture overdispersion in the observations, we suggest basing estimation on a model that is within the class \eqref{MainModel}, but more flexible than the GLM. In particular, we consider the model where the latent variables $\{\alpha_t\}$ are independent and multinomial distributed with $K$ possible outcomes $\left(\alpha_t \in \{S_1,\dots,S_k\}\right)$, and corresponding probabilities ${p_i} = P(\alpha_t = S_i)$, $i = 1,\dots,k$. The process $\{Y_t\}$ is thus treated as following a Poisson mixture model. The likelihood associated with Poisson FMM can be expressed as
\begin{equation} \label{FMMLikelihood}
\mathcal{L} = \prod\limits_{t=1}^n \sum\limits_{i = 1}^K P(Y_t = y_t \mid \alpha_t = S_i) p_i
\end{equation}
where $P$ is the Poisson probability mass function with mean parameter $\mu_t \exp(S_i)$. We define the FMM estimator as the maximizer of \eqref{FMMLikelihood}. For the purpose of simplicity, we restrict our study to the case where $K = 2$. Optimization of the likelihood function is easy in this case and the estimator is of dimension just $d + 2$ for $K = 2$.

Like the GLM estimator, the FMM estimator may not use all of the information in the autocorrelation in the observations. However, it can capture the information in the overdispersion. Therefore, we expect it to be more efficient than the GLM estimator.

\subsection{Poisson HMM}
The third model we consider for estimation purposes is the Poisson HMM, which allows for both autocorrelation and overdispersion. A stationary Poisson HMM can be defined by  allowing ${\alpha_t}$ in \eqref{MainModel} to follow a Markov chain (MC) with $K$ hidden states $\left(\alpha_t \in \{S_1,\dots,S_K\}\right)$, transition probabilities $P_{ij}$, $i,j = 1,\dots,K$, and limiting probabilities $\pi(S_j)$, $j = 1,\dots,K$. As an aside, the Poisson FMM is a special case of the Poisson HMM where the rows of the transition probability matrix, $\{P_{i,j}\}$, are identical.

The likelihood function of the Poisson HMM can be expressed as
\begin{eqnarray}
\label{HMMLikelihood}
\mathcal{L} &=& \sum\nolimits_{\mathbf{\alpha}} \prod\limits_{t=1}^n P(Y_t = y_t \mid \alpha_t) \pi(\alpha_1) \prod\limits_{t=2}^n P_{\alpha_{t-1}, \alpha_t} \\ \nonumber
 &=& \sum\nolimits_{\alpha_1} \pi(\alpha_1) P(Y_1 = y_1 \mid \alpha_1) \sum\nolimits_{\alpha_2} P_{\alpha_1, \alpha_2} P(Y_2 = y_2 \mid \alpha_2) \dots \sum\nolimits_{\alpha_n} P_{\alpha_{n-1}, \alpha_n} P(Y_n = y_n \mid \alpha_n)
\end{eqnarray}
This expression is equivalent to a product of matrices. We define the HMM estimator as the maximizer of \eqref{HMMLikelihood}. For a small number of hidden states, $K$, the HMM estimator is easy to compute using numerical methods. As in the case of the FMM estimator, we thus restrict attention to the case where $K = 2$. We expect the HMM estimator, which can use the information in the overdispersion and autocorrelation in the data but has dimension of just $d + 3$, to be more efficient than the GLM and FMM estimators.

\subsection{Poisson GLMM}
Another estimator considered in the literature is the GLMM estimator (\citealt{Nelson:Leroux:2008}). Specifically, let $\{\alpha_t\}$ be an unobserved $AR(1)$ process and set $\alpha_t = c + \phi \alpha_{t - 1} + \delta_t, t = 1, \dots, n$, where $\delta \sim N(0 , \sigma^2)$. To satisfy the constraint $E(\exp(\alpha_t)) = 1$, set $c = - \frac{\sigma^2}{2 (1 + \phi)}$. Consequently, $\alpha_t \sim N(- \frac{\sigma^2}{2 (1 - \phi^2)}, \frac{\sigma^2}{1 - \phi^2})$ (DDW). Then, the process $\{Y_t\}$ is treated as following a Poisson GLMM. The corresponding likelihood is
\begin{equation}
\label{GLMMLikelihood}
\mathcal{L} = \int_{\alpha_1} \dots \int_{\alpha_n} \prod\limits_{t=1}^n f(y_t \mid \alpha_t) g(\alpha_1, \dots, \alpha_n) d\alpha_1 \dots d\alpha_n
\end{equation}
We define the GLMM estimator as the maximizer of \eqref{GLMMLikelihood}. Like the HMM estimator, it can incorporate information in the overdispersion and autocorrelation. However, unlike \eqref{HMMLikelihood}, \eqref{GLMMLikelihood} does not have a simple algebraic form, and numerical methods typically do not perform well for such high dimensional integrals. \cite{Nelson:Leroux:2008} use Markov chain Monte Carlo methods to approximate the likelihood and obtain the maximizer. Their simulation studies suggest that, when the true model is a Poisson GLMM, the GLMM estimator (i.e. the MLE) is consistent for the regression parameters. However, computations are expensive, particularly for large sample sizes. Thus, we do not explore the properties of the GLMM estimator further in this paper.

\section{SE of the estimators}
\label{estimate SE}

In this section, we develop approximate SEs for the estimators. \cite{White:1984} develops asymptotic covariance matrices for estimators based on dynamic models (which include Poisson HMMs, FMMs and GLMs) even when the models are misspecified. Let $M$ be the (possibly misspecified) model that we fit to the data, and $\theta$ be the vector of parameters in this model. \cite{White:1984} shows that asymptotically $\sqrt{n} I_n ^{* -1/2} H_n ^* (\widehat{\theta_n} - \theta^*) \sim N(0, 1)$, where
\begin{eqnarray}
\label{WhiteSE}
H_n^* &=& E \Bigg[ n ^{-1} \sum\limits_{t=1}^n  \frac{\partial^2 \log g\left( Y_t | Y_1, \dots, Y_{t-1}, X, \theta_n^* \right)}{\partial {\theta_n^*} ^2} \Bigg] \nonumber\\
\mbox{and} \nonumber\\
I_n^* &=& var \Bigg[ n ^{-1/2} \sum\limits_{t=1}^n  \frac{\partial \log g\left( Y_t | Y_1, \dots, Y_{t-1}, X, \theta_n^* \right)}{\partial \theta_n^*} \Bigg].
\end{eqnarray}
Under mild conditions, \cite{White:1984} shows that $\widehat{H_n}$ and $\widehat{I_n}$ are consistent estimators of $H_n$ and $I_n$, where
\begin{eqnarray}
\label{WhiteSE}
\widehat{H_n} &=& \left. n ^{-1} \sum\limits_{t=1}^n \frac{\partial^2 \log g\left( Y_t | Y_1, \dots, Y_{t-1}, X, \theta_n \right)}{\partial \theta_n} \right|_{\theta_n = \widehat{\theta_n}} \nonumber\\
\mbox{and} \nonumber\\
\widehat{I_n} &=& n ^{-1} \sum\limits_{t=1}^n  \frac{\partial \log g\left( Y_t | Y_1, \dots, Y_{t-1}, X, \theta_n \right)}{\partial \theta_n} \frac{\partial \log g\left( Y_t | Y_1, \dots, Y_{t-1}, X, \theta_n \right) ^ t}{\partial \theta_n} \nonumber\\
&+& n ^{-1} \sum\limits_{\tau=1}^{\ell} \sum\limits_{t=\tau + 1}^n  \left\{ \frac{\partial \log g\left( Y_t | Y_1, \dots, Y_{t-1}, X, \theta_n \right)}{\partial \theta_n} \frac{\partial \log g\left( Y_{t - \tau} | Y_1, \dots, Y_{t - \tau - 1}, X, \theta_n \right) ^ t}{\partial \theta_n} \right. \nonumber\\
&+& \left. \left. \frac{\partial \log g\left( Y_t | Y_1, \dots, Y_{t-1}, X, \theta_n \right) ^ t}{\partial \theta_n} \frac{\partial \log g\left( Y_{t - \tau} | Y_1, \dots, Y_{t - \tau - 1}, X, \theta_n \right)}{\partial \theta_n} \right\} \right|_{\theta_n = \widehat{\theta_n}} .
\end{eqnarray}
Here, $\widehat{\theta_n}$ is the MLE of $\theta$ based on the assumed model $M$ and $g$ is the conditional distribution of the observation at time $t$ given previous observations based on $M$. \cite{White:1984} suggests using $\ell < n ^ {1/3}$.

To approximate the SE of the HMM estimator, we use the algorithm of \cite{Lystig:Hughes:2002} to compute partial derivatives of the conditional pmf, $g$, in $\widehat{H_n}$ and $\widehat{I_n}$ efficiently ($g$ is equivalent to $\Lambda_t = P(Y_t | Y_1, \dots, Y_{t - 1})$ in the notation of \cite{Lystig:Hughes:2002}).

For the GLM estimator, DDW propose a SE that depends on well-behaved estimates of the latent process' moments, which are not easy to obtain. As an alternative, since the GLM is a specific case of the HMM, we can use \eqref{WhiteSE} to obtain a SE of the GLM estimator that is independent of the latent process' parameters. Evaluating \eqref{WhiteSE} in the GLM case is straightforward since $\widehat{H_n}$ is the usual estimated covariance matrix of the GLM estimator when the GLM is the true model, and the elements of $\widehat{I_n}$ are products of the first derivatives of the Poisson GLM log-likelihood function.


Equation \ref{WhiteSE} can in fact be used to obtain SEs for all of our estimators, since they are all special cases of the HMM.

\section{Performance of the estimators and their standard errors for special cases of the true model}
\label{Sample SD}
This section details a simulation study of the three estimators described in \S \ref{Model specification}: the GLM, FMM, and HMM estimators. We consider certain true models in the class \eqref{MainModel} to illustrate our key points, and report our general results concerning models in this class in \S \ref{Morestudies}.   

The true models considered in this secion include GLMMs and HMMs, with parameters chosen to achieve different degrees of variation and autocorrelation in the observations (see \eqref{Var} and \eqref{Corr}). Specifically, we consider values of $\sigma_{\alpha}^2$ that lead to a variety of overdispersion (OD) factors, defined as
\begin{eqnarray}
OD_t &\equiv& \frac{Var(Y_t)}{E(Y_t)} \\ \nonumber
&=& 1 + \sigma_{\alpha}^2 \mu_t
\end{eqnarray}
Likewise, we choose a variety of values for $\rho_Y(t)$ (emphasizing in particular $\rho_Y(1)$, which we call AC1) in order to investigate the impact of the autocorrelation on the performance of our estimators.

\begin{figure}
\centering\includegraphics[scale=0.15]{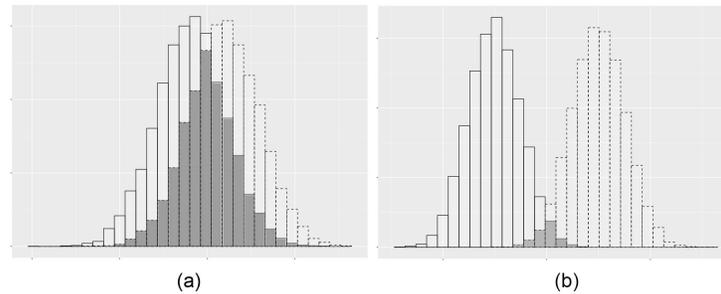}
\caption{Two distributions with low (a) and high (b) separation probability (shaded area).}
\label{Overlap}
\end{figure}

In addition, we consider a third property of the data, which we call ``separation probability" (SP). Specifically, in the case where the latent variable takes on $K$ different values, we have $K$ different conditional distributions for the observations. For each $t$, we define $SP_{S_j, S_{j + 1}}$ as one minus the overlap probability of each ``adjacent" pair of conditional distributions, i.e., if $S_1 < \dots < S_k$, we have the following $k - 1$ values of $SP$:
\begin{eqnarray}
SP_{S_j, S_{j+1}} = 1 - \sum\limits_{i=0}^\infty \min \{ P\left(Y_t = i \mid \alpha_t = S_j \right) , P\left(Y_t = i \mid \alpha_t = S_{j+1} \right) \}
\end{eqnarray}
$j = 1, \dots, k-1$. $SP$ represents the closeness of the conditional Poisson distributions. Figure~\ref{Overlap} shows two examples with low (a) and high (b) SP. In the case where the latent variable is continuous (e.g., in the Poisson GLMM in \eqref{GLMMLikelihood}), we define $SP = 0$.

In summary, we consider three data properties (OD, AC1 and SP) as factors that may affect the efficiency of our estimators. In our simulation studies, we choose the parameters of the true models to achieve different levels of these factors, as described in table~\ref{FactorLevels}. Note that the properties cannot always be manipulated separately. For instance, in the Poisson HMM with $K = 2$ and a fixed transition probability matrix, all the three data properties are increasing functions of the first hidden state, $S_1$.
\begin{table}[!p]
\begin{threeparttable}
\caption{Factor levels in simulation study}
\label{FactorLevels}
\centering
\begin{tabular}{l@{\hskip 0.5in}c@{\hskip 0.5in}c@{\hskip 0.5in}c@{\hskip 0.5in}}
& Low & Medium & High \\[5pt]
OD\tnote{*}  & 1.5 & 3 & 5 \\
AC1\tnote{*} & 0.15 & 0.25 & 0.5 \\
SP\tnote{*+} & 0.25 & 0.45 & 0.7 \\
\end{tabular}
\begin{tablenotes}
      \small
      \item[*]  Averaged over the covariate values
      \item[+] This factor is $0$ for GLMMs
\end{tablenotes}
\end{threeparttable}
\end{table}

We focus on two main cases. In study 1, the true models are chosen such that the MLEs are easy to compute. We then compare the bias and sample variances (SVs) of our estimators to those of the MLE of the true models. In this way, we get a sense of how much information we lose by fitting a misspecified model (under the assumption that the MLE is the most efficient estimator). In study 2, the MLE of the true models are too expensive to compute. Therefore, we simply compare the bias and SVs of our estimators (which are all based on misspecified models).

In our studies, we consider covariates of different forms, including binary, normally distributed, and a trend. We investigate two sample sizes, $n = 100$ and $n = 1000$, and we generate 4000 replicates for each run.

\subsection{Study 1}
We first consider the case where the true model is a stationary Poisson HMM with $K = 2$ hidden states. The 2-state HMM normally has 3 free parameters (say $S_1$, $p_{11}$ and $p_{22}$). However in this study, for simplicity, we fix the elements of the transition probability matrix at $p_{11} = p_{22} = 0.9$ and vary only the value of $S_1$. We use a binary covariate.

\begin{figure}
\centering\includegraphics[width=10cm]{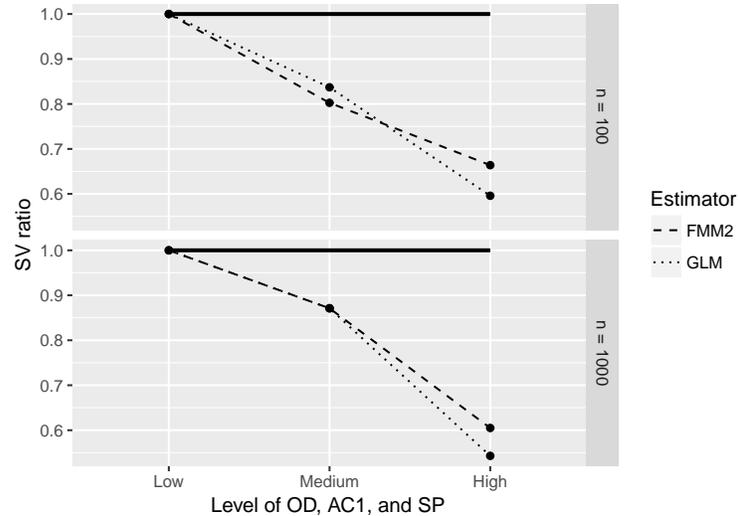}
\caption{Ratio of the SVs of the HMM estimator of the slope (the MLE) to that of the GLM and FMM estimators when a 2-state Poisson HMM is the true model. The solid black line indicates where the efficiency of the estimators would be equal to that of the MLE of the true model. All three factors are simultaneously set to the levels indicated on the x-axis.}
\label{HMM2SER}
\end{figure}

The simulation results show that all the estimators are approximately unbiased even for $n = 100$; the magnitude of the estimated bias was always less than or equal to 0.005 (see online material, table SM 2). Figure~\ref{HMM2SER} shows the ratios of the SVs of the HMM estimator (the MLE of the true model, in this case) to those of the GLM and FMM estimators for different sample sizes and different levels of the factors. The GLM and FMM estimators perform well at the low level of the factors where the true model is close to a GLM. Otherwise, the HMM estimator outperforms the GLM and FMM estimators. The efficiency of the GLM and FMM estimators relative to the HMM estimator is lower at the high level of the factors, where relative efficiency can be as low as $55 \%$. However, in practice, we expect most levels of the factors to be less than our high level. The trends in SV ratios for different sample sizes are similar.

\subsection{Study 2}
In this study, we consider Poisson HMMs with $K = 3$ and $K = 4$ and also a Poisson GLMM as data generating mechanisms. The 3-state and 4-state HMMs have $d + 8$ and $d + 15$ free parameters, respectively. As in study 1, we vary the $S_j$'s and fix the elements of the transition probability matrices at $p_{ii} = 0.9$ and $p_{ij} = 0.1 / (K - 1)$, where $i \neq j$. We use different forms of covariates, including binary, seasonal, and trend, in this study.

We choose the values of the parameters in the true models to achieve different levels of the factors described in Table~\ref{FactorLevels} (except that SP is fixed at $0$ when the Poisson GLMM is the true model).

\begin{figure}
\centering\includegraphics[width=10cm]{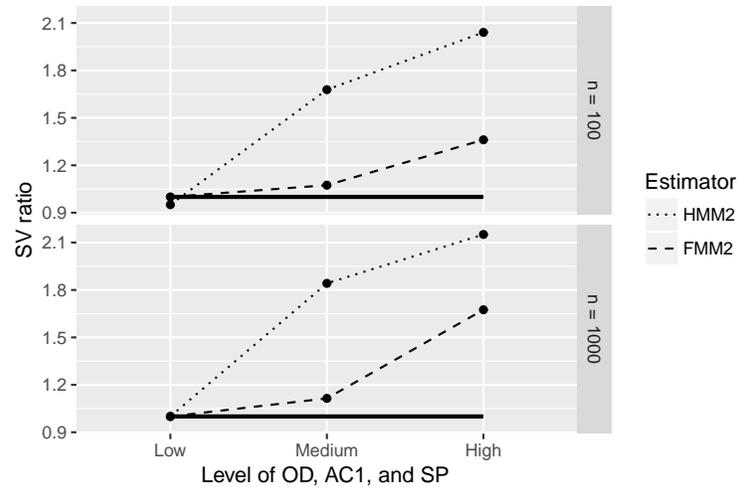}
\caption{Ratio of the SVs of the GLM estimator to those of the HMM and FMM estimators when the 4-state Poisson HMM is the true model. The solid black line indicates where the efficiency of the estimators would be equal to that of the GLM. All three factors are simultaneously set to the levels indicated on the x-axis.}
\label{HMM4SER}
\end{figure}

Our results show that all the estimators are approximately unbiased; the magnitude of the estimated bias was always less than or equal to 0.005 (see online materials, tables SM 3-6 and 8-12). When the Poisson HMM is the true model but the number of hidden states is misspecified, again the GLM and FMM estimators perform well at the low level of the factors. The HMM estimator outperforms all the other estimators at higher levels of the factors (see fig.~\ref{HMM4SER}) and the FMM estimator has the second lowest SV. The results are similar when the true model is a 3-state Poisson HMM (see online supplementary material, table SM 3).

\begin{figure}
\centering\includegraphics[width=10cm]{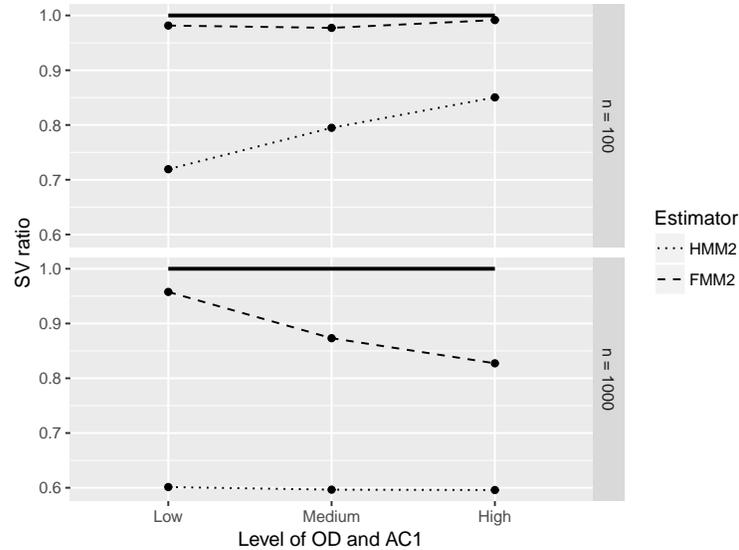}
\caption{Ratio of the SVs of the GLM estimator of the slope to those of the HMM and FMM estimators when the Poisson GLMM is the true model. The solid black line indicates where the efficiency of the estimators would be equal to that of the GLM. OD and AC1 are simultaneously set to the levels indicated on the x-axis. SP is treated as 0.}
\label{AR1SER}
\end{figure}

In contrast, when the true model is a Poisson GLMM (especially with a trend covariate), the efficiency of the GLM estimator is at least as good as those of the HMM and FMM estimators (fig.~\ref{AR1SER}).

We also looked at the performance of the 3-state Poisson HMM and FMM estimators in our simulation study. However, because these estimators were poorly behaved (even for $n = 1000$), we have omitted the details of these findings.

\subsection{Sample SD and SE of the estimators}

In this section, we evaluate our methods in \S\ref{estimate SE} with a simulation study. We treat the sample SDs (SSDs) of the estimates (across replicates) as the true SDs for the purpose of this section, and then compare the SSDs to the average SEs of the estimators. Since we never recommend the FMM estimator (even when the Poisson FMM is the true model, see online materials, table SM 1), we focus only on the SE of the GLM and HMM estimators. In addition, we choose $\ell = 1$ in \eqref{WhiteSE}, after having experimented with different values.

Table~\ref{SEsimul} shows the SSD and the average SE of the GLM and HMM estimators of $\beta_1$ for the high level of the factors, $n = 1000$, and one covariate (binary or trend). The results are similar for other more moderate levels of the factors and for $n = 100$ (see online material, tables SM 21-22). In addition, we compare the average values of the DDW and \cite{White:1984} SEs of the GLM estimator. Both SEs are quite accurate when we have one binary covariate in true model, regardless of the latent process distribution. When the true model has a trend, both underestimate the true SE, but our SE based on the method of \cite{White:1984} always performs better than that of DDW, sometimes dramatically (See online material for other cases, tables SM 21-22, 25-26, and 28).
\begin{table}
\caption{Performance of the SEs of the estimators based on different (misspecified) models}
\label{SEsimul}
{\tabcolsep=4.25pt
\begin{tabular}{lccc|ccc}
 & & \multicolumn{2}{c}{HMM Estimator} & \multicolumn{3}{c}{GLM Estimator} \\[5pt]
True model & Covariate & Sample SD & Asy SE (White)  & Sample SD & Asy SE (DDW) & Asy SE (White)\\[5pt]
 \midrule
Poisson 2-state HMM & Binary & 0.012 & 0.012 & 0.025 & 0.025 & 0.026\\
Poisson 2-state HMM & Trend & 0.012 & 0.012 & 0.068 & 0.034 & 0.039\\
Poisson GLMM  & Binary &  0.030 & 0.030 & 0.025 & 0.024 & 0.025\\
Poisson GLMM & Trend &  0.054 & 0.046 & 0.040 & 0.019 & 0.033\\
\end{tabular}}
\end{table}

\section{General results concerning the estimators and their standard errors}
\label{Morestudies}
Our simulation studies in the previous section illustrate the effect of three factors (OD, AC1 and SP) on the performance of the HMM, FMM and GLM estimators and their SEs for certain true models that have a single covariate. In this section, we report on the performance these estimators more generally, i.e., for any choice of true model in class~\eqref{MainModel}. We study the effect of including multiple covariates as well as the effect of the level of OD, AC1, and SP. With respect to the latter, of the 27 possible combinations of levels of these factors, we study 22. (See online material for details of each setting, table SM 16.) Due to constraints inherent to models in class \eqref{MainModel}, we are not able to simulate data for five combinations, e.g. low OD and high AC1 and SP. We call each combination of factor levels a run.

Table \ref{OverEff} shows a summary of the most efficient estimator (indicated by $*$) for each run where the model had a single covariate. All the estimators are approximately unbiased. In this table, we include only the most extreme runs where the GLM estimator is the most efficient estimator (since models associated with the less extreme runs are even closer to the GLM), and the least extreme runs where the HMM estimator is the most efficient estimator (since models associated with more extreme runs are even closer to the HMM). See online material (table SM 16) for results from other runs. In general, we recommend the GLM estimator as a consistent, efficient and robust estimator. The HMM estimator performs better only if the SP factor is at its high level, or the SP and AC1 factors are both at at least their medium level. In other words, when the data arise from model \eqref{MainModel} with a latent variable that is either continuous or takes on closely spaced values, we recommend the GLM estimator. The HMM estimator is appropriate only when the latent variable takes on highly separated values. We never recommend the FMM estimator. Interestingly, the OD factor doesn't have a big impact on the efficiency of the estimators.
\begin{table}
\caption{Most efficient estimators for different levels of the factors when the model has one covariate}
\label{OverEff}
{\tabcolsep=4.25pt
\begin{tabular}{c@{\hskip 0.45in}cccc@{\hskip 0.35in}ccc}
 & \multicolumn{3}{c}{Factor} & & \multicolumn{2}{c}{Estimator} \\[5pt]
Run & OD & AC1 & SP && GLM & HMM2 \\[5pt]
 \midrule
1 &  High & Medium & Low && * & \\
2 &  Medium & High & Low && * & \\
3 &  High & Low & Medium && * & \\
4 &  Low & Medium & Medium &&  & *\\
5 &  Medium & High & Medium &&  & *\\
6 &  Medium & Low & High &&  & *\\
\end{tabular}}
\end{table}

We also conducted studies where the true model had multiple covariates of different forms. Our results show that the number and form of covariates affect the efficiency of the estimators (see online materials, tables SM 17-28). For instance, estimating a trend effect is challenging and the chance of underestimating its SE is high when using either the HMM or GLM estimator, especially when other covariates are present in the model. In general, the SE of the GLM estimator is better-behaved than that of the HMM estimator when we have many covariates in the model. In particular, when the HMM estimator is more efficient (e.g. runs 4-6 in table \ref{OverEff}), the SE of the HMM estimator can be severely negatively biased (as low as $50\%$ of the SSD), whereas the SE of the GLM estimator is approximately unbiased (except the estimator of the coefficient of the trend, where both SEs are negatively biased). Therefore, we recommend the GLM estimator, regardless of the levels of the factors in table \ref{FactorLevels}.

In practice, when the response is a function of one covariate, the factors in table \ref{FactorLevels} can be estimated. Thus, the problem can be classified according to the runs in table \ref{OverEff} and the best estimator identified. However, when we have multiple covariates in the model, estimating factors described in table \ref{FactorLevels} could be challenging. As an alternative, we can fit a GLM to the observed counts, compute the standardized residuals, and then estimate the factors for these residuals. In this way, we can classify models with any number of covariates according to table\ref{OverEff}.

To summarize, we present some guidelines for choosing the ``best" estimator (i.e., the estimator with high relative efficiency and well-behaved SE) among the three estimators in practice (where the true model is unknown). In particular, we recommend the HMM estimator only when the model has fewer than 4 covariates and SP is at its high level (or SP and AC1 are both at their medium level). Otherwise, in light of its consistency, efficiency, robustness, and well-behaved SE estimate, we recommend the GLM estimator.

\section{Application}
\label{application}

We now apply our findings from the previous sections to real applications. We first re-analyze the polio data considered by \cite{Zeger:1988} and DDW. We then present a second application concerning the daily numbers of epileptic seizures.

\subsection{Polio Incidence Data}
\begin{figure}
\centering\includegraphics[width=10cm]{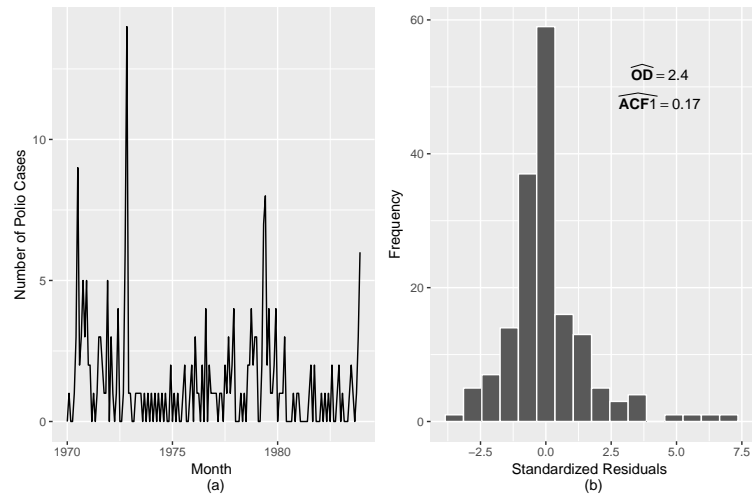}
\caption{(a) Time series plot of polio data and (b) histogram of standardized residuals obtained after fitting the Poisson GLM}
\label{Polio}
\end{figure}

\cite{Zeger:1988} analyzed the monthly number of cases of polio reported by the U.S. Centers for Disease Control from 1970 - 1983 (168 months) to assess whether the data provided evidence of a long-term decrease in the rate of infection. His analysis was based on a QMLE of the time trend.
The data are displayed in fig.~\ref{Polio} (a). The polio counts display overdispersion relative to the Poisson distribution ($\widehat{OD} = 2.40$). To compare our approach with that of DDW, we consider the same set of covariates.
No separation among latent distribution is visible (see the Poisson GLM residuals in fig.~\ref{Polio} (b)), estimated AC1 and OD of the Poisson GLM standardized residuals are low (0.17) and medium (2.40), respectively, suggesting that the true data generating mechanism is less extreme that that in run 1 in table \ref{OverEff}. In addition, we have multiple covariates, including a trend. Therefore, the GLM estimator is the best choice in terms of efficiency and accuracy of its SE.

Table~\ref{Poliofit} shows the GLM estimates of the regression parameters along with their SEs based on our approach and on that of DDW. The SEs based on the two methods are considerably different for all the covariate coefficient estimates (especially the trend). DDW reported the SEs of the GLM estimates (second column in table~\ref{Poliofit}) based on some additional model assumptions (including the assumption that the latent process follows an AR(1) model; see \S4 of \citealt{Zeger:1988}). They didn't offer a way to check these assumptions and we can't verify them easily. Without these assumptions, DDW's SEs of the GLM estimates can be computed using method of moment estimates of the latent process covariances (see DDW and \citealt{Zeger:1988}). We compute the SE of DDW based no further assumptions about the true model (listed in the third column of table~\ref{Poliofit}), which differ from the values reported in Table 1 of DDW and are quite similar to the SEs of the GLM estimates. We recommend using our SE based on the approach of \cite{White:1984} (listed in the fourth column of table~\ref{Poliofit}) because it performs better in general (according to our simulation study) . Nevertheless, the trend is not significant based on either our SE estimate or that of DDW.
\begin{table}
\caption{GLM estimates and SEs (polio dataset). DDWs' SE is computed using the MOM estimators of the latent process parameters (see \S3.2 of DDW). They differ from the values reported in Table 1 of DDW, which rely on additional model assumptions (see \S4 of \citealt{Zeger:1988}).}
\label{Poliofit}
{\tabcolsep=4.25pt
\begin{tabular}{l@{\hskip 0.3in}cccc}
Covariate & Est & Asy SE (DDW - extra assumptions) & Asy SE (DDW) & Asy SE (White) \\[5pt]
\midrule
Intercept & 0.207 & 0.205 & 0.040 & 0.112 \\
Trend $\times 10 ^ {-3}$ & -4.799 & 4.115 & 1.788 & 2.548 \\
$cos(2 \pi t / 12)$ & -0.149 & 0.157 & 0.009 & 0.136 \\
$sin(2 \pi t / 12)$ & -0.532 & 0.168 & 0.082 & 0.191 \\
$cos(4 \pi t / 12)$ & 0.169 & 0.122 & 0.035 & 0.149 \\
$sin(4 \pi t / 12)$ & -0.432 & 0.125 & 0.046 & 0.149\\
\end{tabular}}
\end{table}

\subsection{Daily number of epileptic seizures}

The second application that we consider is a series of counts of myoclonic seizures suffered by one patient on 204 consecutive days. In the neurology literature, Poisson HMMs appear to be common for the analysis of seizure counts (see e.g. Hopkins et al 1985 or Franke and Seligmann 1993). In particular, \cite{{Albert:1991}} and \cite{Le:Leroux:Puterman:1992} fit a Poisson 2-state HMM to these counts. Since other predictor variables for this patient are not available, we consider only a trend effect in the model.

\begin{figure}
\centering
\includegraphics[width=10cm]{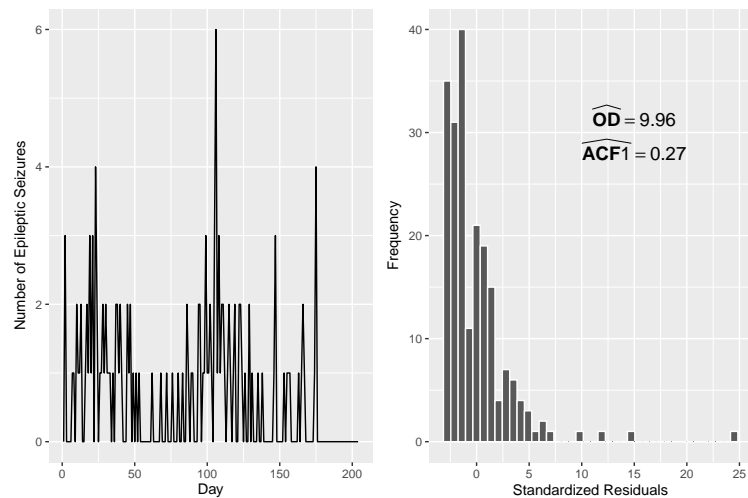}
\caption{(a) Time series plot of epileptic seizure counts and (b) histogram of standardized residuals obtained after fitting the Poisson GLM}
\label{Seizure}
\end{figure}

The data are illustrated in fig.~\ref{Seizure} (a). We can see medium estimated AC1 and high estimated OD in the GLM residuals. In addition, the histogram of the Poisson GLM standardized residuals suggests at least two components (i.e., the true model might be a Poisson HMM), but with low SP (fig.~\ref{Seizure} (b)). Since the properties of the residuals are similar to those of run 1 in table \ref{OverEff}, we recommend the GLM estimator.
Table~\ref{seizurefit} shows the estimated coefficients and their SEs. As expected, the GLM estimator does have smaller SEs than does the HMM estimator.
\begin{table}
\caption{GLM and HMM estimates and SEs (seizure dataset).}
\label{seizurefit}
{\tabcolsep=4.25pt
\begin{tabular}{lccccc}
 & \multicolumn{3}{c}{GLM Estimate}{\hskip 0.3in} & \multicolumn{2}{c}{HMM Estimate}{\hskip 0.3in} \\[5pt]
Covariate & Est & Asy SE (DDW) & Asy SE (White) & Est & Asy SE (White)\\[5pt]
\midrule
Intercept & 0.179 & 0.481 & 0.42 & -0.170 & 0.283\\
Day & -0.006 & 0.363 & 0.172 & -0.933 & 0.820\\
\end{tabular}}
\end{table}

These results are consistent with our findings that the HMM estimator is more efficient only when the true model is an ``extreme" HMM.

\section{Discussion}
\label{discussion}

In this paper, we considered a general class of models for time series of counts \eqref{MainModel}. We conducted a comprehensive study of the accuracy and efficiency of three estimators, the GLM, 2-state FMM and 2-state HMM estimators, and the accuracy of their SEs.

Our results showed that except in extreme cases, the GLM estimator is the most efficient.  In addition, the GLM estimator is consistent and robust, has a well-behaved estimated SE, and is easy to compute using standard software. The GLM estimator has particular advantages over the MLE of the Poisson GLMM. Specifically, the Poisson GLMM allows for the latent variable to have a general (i.e. possibly non-normal) distribution. However, we usually don't know the distribution of the latent variable. This uncertainty provides yet more motivation for using the GLM estimator.

We considered other estimators including the 3-state FMM and HMM estimators, negative binomial (\citealt{Davis:Wu:2009}), and Zeger QMLE, as well, but they were less efficient than the 2-state HMM estimator, or poorly behaved, or their SEs were poorly behaved (see \S\ref{Sample SD}).

Initially, we thought of the HMM with $x$ number of hidden states (HMMx) as approximation to the GLMM. Thus, we expected that the HMMx estimator (for $x > 2$) would perform better than the GLM or HMM2 estimator when the true model was a GLMM, HMM3, HMM4, etc. But that was not the case, at least for the sample sizes we considered. Interesting future work could include the exploration of complex models where the HMMx estimator does perform well.

We proposed three main factors (OD, AC1 and SP) associated with the true model that could affect the performance of the estimators. We then considered a broad set of simulation runs (including extreme cases) by changing the levels of our factors. Surprisingly, OD, which is an obvious violation of the assumption of a Poisson GLM, had limited effect on the efficiency of the estimators -- even the Poisson GLM estimator.

In addition, we developed SEs for our estimators. To the best of our knowledge, our SEs for the Poisson HMM and FMM estimators are unique, our SE for the Poisson GLM estimator is easier to compute and more accurate than the existing SEs.

We focused on the efficiency of the covariate coefficient estimators. However, in our simulation studies, we also considered the intercept estimators. Our results showed that all of our intercept estimators are approximately unbiased. In addition, both DDW's and our SEs underestimate the true SD of both the GLM and HMM estimators. But in general, our SE was relatively closer to the SSD of the GLM intercept estimates. Developing a SE that is (approximately) unbiased for all the coefficient estimators (including the intercept), is additional future work.

\CJShistory
\end{document}